\documentclass[sigconf]{acmart}
\acmConference[EASE 2025]{The 29th International Conference on Evaluation and Assessment in Software Engineering}{17–20 June, 2025}{Istanbul, Turkey}
\usepackage[T1]{fontenc}
\usepackage{graphicx} 
\usepackage{paralist}
\usepackage{colortbl}
\usepackage[skip=0.5\baselineskip,justification=centering]{caption}
\usepackage{subcaption}
\usepackage{formal-grammar}
\usepackage[capitalise, noabbrev]{cleveref}
\Crefname{lstlisting}{Listing}{Listings}
\usepackage[linesnumbered,ruled]{algorithm2e}
\usepackage{xcolor}
\usepackage{soul}
\usepackage{arydshln}
\usepackage{listings}

\lstdefinestyle{mystyle}{
  language=Java, 
  upquote=true,
  basicstyle=\scriptsize\ttfamily,
  breakatwhitespace=false,         
  breaklines=true,                 
  captionpos=b,                    
  keepspaces=true,                 
  showspaces=false,                
  showstringspaces=false,
  showtabs=false,                  
  tabsize=2,
  escapeinside={(*@}{@*)},
}

\definecolor{lgreen}{RGB}{144, 238, 144}
\definecolor{lred}{RGB}{255, 114, 118}

\title{An Anatomy of 488 Faults from Defects4J Based on the Control- and Data-Flow Graph Representations of Programs}
\author{Alexandra van der Spuy}
\email{23552395@sun.ac.za}
\affiliation{%
  \institution{Stellenbosch University}
  \city{Stellenbosch}
  \country{South Africa}}

\author{Bernd Fischer}
\email{bfischer@sun.ac.za}
\affiliation{%
  \institution{Stellenbosch University}
  \city{Stellenbosch}
  \country{South Africa}}
\date{February 2025}

\begin{document}

\begin{abstract}
Software fault datasets such as Defects4J provide for each individual fault its location and repair, but do not characterize the faults. 
Current classifications use the repairs as proxies, but these do not capture the intrinsic nature of the fault.  
In this paper, we propose a new, direct fault classification scheme based on the control- and data-flow graph representations of programs.  
Our scheme comprises six control-flow and two data-flow fault classes.  
We manually apply this scheme to 488 faults from seven projects in the Defects4J dataset.  
We find that the majority of the faults are assigned between one and three classes. 
We also find that one of the data-flow fault classes (definition fault) is the most common individual class but that the majority of faults are classified with at least one control-flow fault class.  
Our proposed classification can be applied to other fault datasets and can be used to improve fault localization and automated program repair techniques for specific fault classes.
\end{abstract}



\settopmatter{printacmref=false}
\setcopyright{none}
\renewcommand\footnotetextcopyrightpermission[1]{}
\pagestyle{plain}

\maketitle

\section{Introduction}
Software debugging is an expensive and time-consuming process~\cite{DiGiuseppe2015,Defects4JDissection}, requiring faults to be found (\emph{fault localization}) and fixed (\emph{fault repair}).
Its effectiveness depends on several factors, but one factor that is often overlooked is the nature of the fault itself~\cite{Nath_Merkel_Lau_2012}.
Typically, this fault nature is unknown, but an analysis of known real-world faults allows us to identify common fault types, which can in turn be used to improve fault localization and repair techniques.

However, investigating the \emph{intrinsic nature} of faults can be labor-intensive,
and classifications typically use the applied repairs as proxies
\cite{Pan_Kim_Whitehead_2009,Nath_Merkel_Lau_2012,Lucia_Lo_Jiang_Thung_Budi_2014,Defects4JDissection}.
This allows automating the classification process but it does not necessarily
capture the intrinsic nature of the faults themselves, since faults can
typically be repaired in different ways. 
Consider for example \cref{listing:lang7,listing:math46}:
\sethlcolor{lred}
\begin{lstlisting}[caption={Conditional block repair---removal (Lang-7)~\cite{Defects4JDissection}},label={listing:lang7}]
(*@\hl{-}@*) (*@\hl{\textbf{if} (str.startsWith("--")) \{}@*)
(*@\hl{-}@*)   (*@\hl{return \textbf{null};}@*)
(*@\hl{-}@*) (*@\hl{\}}@*)
\end{lstlisting}
In \cref{listing:lang7}, 
the underlying fault is not the conditional, but the return-statement; removing only the return would also be a valid fix.
\begin{lstlisting}[caption={Unwraps-from repair---if-else (Math-46)~\cite{Defects4JDissection}},label={listing:math46}]
(*@\sethlcolor{lred}\hl{-}@*) return (*@\sethlcolor{lred}\hl{isZero ? }@*)NaN(*@\sethlcolor{lred}\hl{ : INF}@*);
(*@\sethlcolor{lgreen}\hl{+}@*) return NaN;
\end{lstlisting}
For \cref{listing:math46}, 
the repair is classified as \emph{unwraps-from} because the conditional is removed, but the true fault is in referencing and returning \texttt{INF} instead of \texttt{NaN}; another possible repair is to keep the conditional and simply replace \texttt{INF} with \texttt{NaN}.

In this paper, we describe a new fault classification scheme based on the control- and data-flow graph representations of programs, and a corresponding dataset.  
The main novelty of our scheme is its grounding in the (abstract) structural properties of the faults themselves, rather than the applied repairs. 
This emphasizes the commonalities of different faults under syntactic variations. 
Consider for example the faults (and repairs) shown in \cref{listing:lang62,listing:chart18}. 
In both cases, the underlying problem is that the control flow from the fault location to the next location is wrong (i.e., whenever the program  executes the faulty statement, it executes the wrong statement next), despite the differences in the faulty statements (i.e., return vs.\ missing statement) and the repair (i.e., substitution of return by throw vs.\ insertion of break).
We thus classify both faults as unconditional control-flow target faults.

Our scheme comprises six control-flow and two data-flow fault classes.  
We manually classify 488 faults from seven projects in the Defects4J dataset accordingly.
The most common individual class is
one of the data-flow fault classes (definition faults); however, the
control-flow classes are overall more common.  Similar to repair
classes~\cite{Defects4JDissection}, our fault classes are not disjoint since
a single fault can have characteristics of multiple classes.
The majority of the faults are thus assigned one to three of the eight classes.

Our classification is useful for investigating intrinsic characteristics of faults, rather than their repairs.
We currently use it to evaluate the effectiveness of different spectrum-based fault localization (SBFL) techniques over the different fault classes,
and our preliminary results show that SBFL is more effective for pure data-flow than pure control-flow faults.
We also see 
possible applications in automated program repair. For example, we can use the 
fault class frequency information 
to guide automated program repair to attempt fixes in order of fault class frequency; if we can classify individual 
faults on-the-fly, we could even tailor individual repair operations to the specific fault classes.

In this paper, we make two main contributions. 
First, we introduce a novel flow graph-based fault classification scheme, which we present with examples in \Cref{sec:fgbc}. 
Second, we construct a dataset of faults classified according to this scheme. 
We explain our construction methodology in \Cref{sec:dcm} and analyze the dataset statistics in \Cref{sec:ds};
dataset and replication package are publicly available (see \Cref{sec:da}).
We then consider several ways to use the dataset to answer research questions in \Cref{sec:duafrq}.
We summarize related work in \Cref{sec:rw} and our conclusions in \Cref{sec:conclusion}. 

\section{Flow Graph-Based Classification}\label{sec:fgbc}

\subsection{Flow graphs}\label{subsec:fg}
We base our classification on flow graph representations of programs.
Specifically, we consider the control-flow graph (CFG) and data-flow graph
(DFG) as a single structure $G=(V,E_C,E_D)$, where the nodes $V$
correspond to the program statements, the CFG edges $E_C$
correspond to changes in the program counter, and the DFG edges $E_D$
represent associations between variable writes or \emph{defs} and their
references or \emph{uses}.
The flow graph construction, with explicit def-use association edges, is fairly
standard \cite{Allen1970,Ribeiro2019}.  For the classification we assume that
($i$) jumps (i.e., break-, continue-, return-, and
throw-statements) are represented by CFG edges only, and not by
nodes, and ($ii$) method calls are represented by a node modeling the
parameter passing and a CFG edge modeling the jump.


\subsection{Control-flow faults}\label{subsec:cf}

\subsubsection{Unconditional control-flow target faults}\label{subsubsec:unconditionaltargetf}

Such faults occur when an CFG edge has the wrong target
and the source of this faulty edge 
has no other outgoing edge, i.e., does not leave from a predicate node. Any 
run that reaches the source node will thus traverse the faulty edge and inevitably enter an error state, but not necessarily fail.
Faulty edges can correspond to wrong statement orders, or to missing,
extraneous, or wrong jumps and method calls; we thus distinguish three different classes of unconditional control-flow faults, as discussed below.
Furthermore, for classification as an unconditional control-flow fault we require that the flow graph already contains the correct set of nodes,
unless the fault occurs in combination with other faults.

\paragraph{Statement order faults}\label{par:order} 
Such faults occur when a sequence of consecutive edges connect multiple
nodes in the wrong order, but the set of source and target nodes of these
edges is correct; a repair rearranges the edges' sources and targets, which
corresponds at the source level to a reordering of the statements. Closure-13 shows an example:
\begin{lstlisting}[caption={Statement order fault (Closure-13)},label={listing:closure13}]
(*@\sethlcolor{lred}\hl{+}@*) (*@\sethlcolor{lred}\hl{traverse(c);}@*)
  Node next = c.getNext();
(*@\sethlcolor{lgreen}\hl{+}@*) (*@\sethlcolor{lgreen}\hl{traverse(c);}@*)
  c = next; 
\end{lstlisting}

\paragraph{Jump faults}\label{par:jump} 
A jump fault occurs if the target of a single edge is wrong; this edge
can connect two adjacent statements, representing the normal control
flow, or span a larger gap, representing a jump statement.  In both cases the
repair effectively ``re-wires'' this edge.
For example, Chart-18 contains a \emph{wrong} jump, where the
return-statement induces a CFG edge with the wrong target; replacing it with
a throw-statement induces an edge with the correct target:
\begin{lstlisting}[caption={Jump fault---wrong jump target (Chart-18)},label={listing:chart18}]
  if (index < 0) {
(*@\sethlcolor{lred}\hl{-}@*)   (*@\sethlcolor{lred}\hl{\textbf{return};}@*)
(*@\sethlcolor{lgreen}\hl{+}@*)   (*@\sethlcolor{lgreen}\hl{\textbf{throw} new UnknownKeyException("The key (" + key}@*)
(*@\sethlcolor{lgreen}\hl{+}@*)       (*@\sethlcolor{lgreen}\hl{    + ") is not recognised.");}@*)
  }
\end{lstlisting}
In contrast, Lang-62 contains a \emph{missing}
jump fault where the control flow incorrectly connects the assignment
to the next case-label, and the insertion of the break-statement repairs this.
Conversely, removing a faulty \emph{extraneous} jump restores the
normal control flow.
\begin{lstlisting}[caption={Jump fault---missing jump (Lang-62)},label={listing:lang62}]
  case 'x' : {
    entityValue = Integer.parseInt(entityContent
        .substring(2), 16);
(*@\sethlcolor{lgreen}\hl{+}@*)   (*@\sethlcolor{lgreen}\hl{\textbf{break};}@*)
  }
\end{lstlisting}

\paragraph{Method call faults}\label{par:method}
The similarity between jumps and method calls also extends from the CFG structure to faults. The only
small difference is that the \emph{repair} of a missing
method call with arguments also requires inserting the node
that models the parameter passing; this implies that such faults are also
definition faults (\S\ref{subsubsec:def}).
Listings~\ref{listing:time22} and \ref{listing:math64} show a \emph{wrong} resp.\ \emph{missing} method call (cf.\ 
Listings~\ref{listing:chart18} and \ref{listing:lang62} for the corresponding jump faults).
\sethlcolor{lgreen}
\begin{lstlisting}[caption={Method call fault---wrong method (Time-22)},label={listing:time22}]
(*@\sethlcolor{lred}\hl{-}@*) (*@\sethlcolor{lred}\hl{\textbf{this}(duration, \textbf{null}, \textbf{null});}@*)
(*@\sethlcolor{lgreen}\hl{+}@*) (*@\sethlcolor{lgreen}\hl{\textbf{super}();}@*)
\end{lstlisting}
\vspace{-2ex}
\begin{lstlisting}[caption={Method call fault---missing call (Math-64)},label={listing:math64}]
  if (maxCosine <= orthoTolerance) {
    // convergence has been reached
(*@\hl{+}@*)   (*@\hl{updateResidualsAndCost();}@*)
\end{lstlisting}

\subsubsection{Control-flow predicate faults}\label{subsubsec:guardaroundbodyf}
Control-flow predicates (i.e., the conditions of conditionals and loops)
determine which CFG paths a program follows.  A control-flow predicate fault
occurs if at least one of these paths is incorrect.  This can be caused by a
fault in the condition itself, missing/extraneous predicate nodes
guarding otherwise correct existing nodes, or entire
predicate blocks that are missing or superfluous. We therefore distinguish three different classes of control-flow predicate faults, as discussed below.

In contrast to unconditional jump target faults, a program with a control-flow
predicate fault may reach the fault location without entering an erroneous state (e.g.,
a run of the fault entry of Lang-55 shown in \cref{listing:lang55}, in
which the inserted predicate would evaluate to true, does not cause an error).

\paragraph{Predicate node faults}\label{par:pred}
This class captures faults where the CFG structure is correct, but
the label of a predicate node (i.e., the condition) is wrong.  
For example, Chart-1 contains a predicate node
fault, but if the faulty node is also the target of a faulty DFG edge (i.e., 
references a wrong variable), it is additionally classified as a data-flow fault.
\begin{lstlisting}[caption={Predicate node fault (Chart-1)},label={listing:chart1}]
(*@\sethlcolor{lred}\hl{-}@*) if (dataset (*@\sethlcolor{lred}\hl{!=}@*) null) {
(*@\sethlcolor{lgreen}\hl{-}@*) if (dataset (*@\sethlcolor{lgreen}\hl{==}@*) null) {
    return result;
  }
\end{lstlisting}

\paragraph{Predicate existence faults}\label{par:guard}
This class captures faults where the CFG has an
extraneous (resp.\ missing) predicate node guarding existing nodes, leading to paths that execute the guarded nodes in too few (resp.\ many) paths.
For example, Lang-55 contains a \emph{missing} if-guard fault, which causes the existing assignment to \texttt{stopTime} to be executed too often: 
\sethlcolor{lgreen}
\begin{lstlisting}[caption={Predicate existence fault---missing if (Lang-55)},label={listing:lang55}]
  }
(*@\hl{+}@*) (*@\hl{\textbf{if}(\textbf{this}.runningState == STATE\_RUNNING) \{}@*)
    stopTime = System.currentTimeMillis();
(*@\hl{+}@*) (*@\hl{\}}@*)
  this.runningState = STATE_STOPPED;
\end{lstlisting}

\paragraph{Predicate block faults}\label{par:block}
In contrast to predicate \emph{existence} faults, predicate \emph{block} faults
capture situations where the CFG is missing not only the predicate node but also
the entire guarded block.
Closure-85 exhibits such a fault:
\sethlcolor{lgreen}
\begin{lstlisting}[caption={Predicate block fault---missing while (Closure-85)},label={listing:closure85}]
  Node next = ControlFlowAnalysis.computeFollowNode(n);
(*@\hl{+}@*) (*@\hl{\textbf{while} (next != \textbf{null} \&\& next.getType() == Token.BLOCK)}@*)
(*@\hl{+}@*)   (*@\hl{\textbf{if} (next.hasChildren())}@*)
(*@\hl{+}@*)     (*@\hl{next = next.getFirstChild();}@*)
(*@\hl{+}@*)   (*@\hl{\textbf{else}}@*)
(*@\hl{+}@*)     (*@\hl{next = computeFollowing(next);}@*)
  return next; 
\end{lstlisting}

\subsection{Data-flow faults}\label{subsec:df}
Data-flow faults occur when the value transported along a DFG edge does not 
match the requirements of the read at the target of this edge. The edge itself can be correct but the corresponding value is incorrect, or the source of the edge can be wrong, either caused by a faulty or a missing definition, or by a faulty use. 

\subsubsection{Definition faults}\label{subsubsec:def}
Such faults occur at the source of a DFG edge. In the simplest case, the source location is correct, but the value written to the memory location (i.e., variable) used at the target location is wrong. Closure-97 exhibits such a fault;
note that this is not a use fault, despite the variable change from \texttt{lvalInt} to \texttt{lvalLong}, because \texttt{lvalLong} does not exist in the faulty program.

\begin{lstlisting}[caption={Definition fault---wrong write value (Closure-97)},label={listing:closure97}]
(*@\sethlcolor{lgreen}\hl{+}@*) (*@\sethlcolor{lgreen}\hl{\textbf{long} lvalLong = lvalInt \& 0xffffffffL;}@*)
(*@\sethlcolor{lred}\hl{-}@*) result = (*@\sethlcolor{lred}\hl{lvalInt}@*) >>> rvalInt;
(*@\sethlcolor{lgreen}\hl{+}@*) result = (*@\sethlcolor{lgreen}\hl{lvalLong}@*) >>> rvalInt;
\end{lstlisting}

Alternatively, the source location of the edge may be wrong, either because the correct value is written to the wrong location (see \cref{listing:math21}), or because a required write is missing (see \cref{listing:lang17}).

\begin{lstlisting}[caption={Definition fault---wrong location (Math-21)},label={listing:math21}]
  // find maximal diagonal element
(*@\sethlcolor{lred}\hl{-}@*) (*@\sethlcolor{lred}\hl{swap[r]}@*) = r;
(*@\sethlcolor{lgreen}\hl{+}@*) (*@\sethlcolor{lgreen}\hl{\textbf{int} swapR}@*) = r;
  for (int i = r + 1; i < order; ++i) {
\end{lstlisting}

\begin{lstlisting}[caption={Definition fault---missing definition (Lang-17)},label={listing:lang17}]
  out.write(c);                
(*@\sethlcolor{lgreen}\hl{+}@*) (*@\sethlcolor{lgreen}\hl{pos+= c.length;}@*)
\end{lstlisting}

\subsubsection{Use faults}\label{subsubsec:use}
Such faults occur at the target of a DFG edge, where the value is read from the wrong location.
Jsoup-57 contains a faulty use of \texttt{attributes} instead of \texttt{it}:

\begin{lstlisting}[caption={Use fault (Jsoup-57)},label={listing:jsoup57}]
(*@\sethlcolor{lred}\hl{-}@*) (*@\sethlcolor{lred}\hl{attributes}@*).remove((*@\sethlcolor{lred}\hl{attrKey}@*));
(*@\sethlcolor{lgreen}\hl{+}@*) (*@\sethlcolor{lgreen}\hl{it}@*).remove();
\end{lstlisting}

Note that both wrong location definition faults and use faults involve a wrong DFG edge source location; the difference is determined by the corresponding variable: if the variable read is correct, the left-hand side of an assignment needs to be changed to ``rewire'' the edge, otherwise the used variable needs to be changed.
Note also that use faults often occur in conjunction with other faults, e.g., a definition or method call fault.
The latter occurs in \cref{listing:jsoup57} because \texttt{remove(attrKey)} is called instead of \texttt{remove()}.

\section{Dataset Construction Methodology}\label{sec:dcm}
We use Defects4J~\cite{Defects4Jinfo}, a repository of real-world Java projects, as our data source.
Defects4J is often employed to test the effectiveness of fault localization and repair techniques~\cite{Marcilio_2023,Pearson2017,Defects4JDissection}, making it suited for our aim of creating a classification dataset for testing purposes.
We classify all 488~faults from seven Defects4J projects (see \cref{tab:project}) with manual thematic analysis~\cite{Defects4JDissection}.
Our classification can be compared to Sobreira \emph{et al.}~\cite{Defects4JDissection}'s repair classification of 395 of these faults, thus we use Defects4J~v1.2.0~\cite{Defects4J1.2.0} for the relevant projects.
However, we consider another 93~faults of Jsoup, heretofore unclassified, in Defects4J~v3.0.1\cite{Defects4J}.

Each fault is inspected in terms of the source and repaired program, as well as failing test cases.
The fault is then assigned the appropriate fault classes from our classification scheme.
The first author of this paper performed manual classification, which was validated by the second author's review, and the classification was adjusted accordingly.
Disagreements regarding the classification of individual faults were resolved via discussions between both authors.
Manual classification was used to gain an understanding of the intrinsic nature of each fault, allowing us to take into account subtleties that an algorithm may overlook.

\begin{table*}[t!]
  \centering
  \caption{Project characteristics~\cite{Ribeiro2019,Defects4Jinfo,Marcilio_2023}, frequency of fault class assignments, and frequency of distinct fault types}\label{tab:project}
  {\small \tabcolsep4pt \begin{tabular}{|l|r||r|r|r|r|r|r|r|r||r||r|r|r|r|}
    \hline
    & & \multicolumn{9}{c|}{Fault classes} & \multicolumn{4}{c|}{Fault types}\\
    \cline{3-15}
    & & order & jump & call & pred & guard & block 
    & def & use
    & 
    & & & &
    \\
    Project & KLOC & \S\ref{par:order} & \S\ref{par:jump} & \S\ref{par:method} & \S\ref{par:pred} & \S\ref{par:guard} & \S\ref{par:block} & \S\ref{subsubsec:def} & \S\ref{subsubsec:use}
    & Avg per fault & CF & DF & CF/DF & All\\
    \hline
    Chart & $96$ 
    & $1$ & $3$ & $5$ & $4$ & $11$ & $3$ 
    & ${15}$ & $7$ 
    & $1.88\pm 1.15$
    & $11$ & $6$ & $9$ & $26$
    \\
    Closure & $90$ 
    & $6$ & $9$ & $37$ & ${42}$ & $53$ & $30$
    & ${81}$ & $6$ 
    & $1.98\pm 1.03$
    & $52$ & $14$ & $67$ & $133$
    \\
    Lang & $22$ 
    & $2$ & $18$ & $11$ & $11$ & $31$ & $11$ 
    & ${43}$ & $8$ 
    & $2.08\pm 0.86$
    & $20$ & $12$ & $33$ & $65$
    \\
    Math & $85$ 
    & $1$ & $17$ & $29$ & $24$ & $36$ & $18$ 
    & ${83}$ & $20$ 
    & $2.15\pm 1.04$
    & $22$ & $27$ & $57$ & $106$
    \\
    Mockito & $11$ 
    & $0$ & $3$ & $15$ & $2$ & $16$ & $6$ 
    & ${28}$ & $4$ 
    & $1.95\pm 0.86$
    & $10$ & $4$ & $24$ & $38$
    \\
    Time & $28$ 
    & $3$ & $9$ & $6$ & ${7}$ & $17$ & $3$ 
    & ${21}$ & $5$ 
    & $2.63\pm 1.61$
    & $6$ & $4$ & $17$ & $27$\\
    \hdashline
    Jsoup & $10$ 
    & $3$ & $4$ & $36$ & $14$ & $30$ & $23$ 
    & ${59}$ & $6$ 
    & $1.88\pm 0.85$
    & $33$ & $14$ & $46$ & $93$
    \\
    \hline
    All & $342$ 
    & $16$ & $63$ & $139$ & $104$ & $194$ & $94$ 
    & ${330}$ & $56$ 
    & $2.04\pm 1.03$
    & $154$ & $81$ & $253$ & $488$
    \\
    \hline
  \end{tabular}}
  \vspace{-1ex}
\end{table*}

We consider the difference in the flow graphs of the faulty and fixed code to manually assign fault classes according to the intrinsic fault nature.
We infer the nature of the fault from the program context, taking into account 
the intended program semantics expressed by the repair.
However, we investigate alternative repairs and compare all repair flow graphs to the faulty flow graph---the applied repair does not drive our classification.

\paragraph{Limitations}
We consider, similarly to Sobreira \emph{et al.}~\cite{Defects4JDissection}, each Defects4J fault entry as a single fault that may display characteristics of multiple fault classes.
However, certain entries do contain multiple independent faults (e.g., Chart-18).
For comparability and compatibility with the repair classification~\cite{Defects4JDissection}, we do not distinguish between the multiple faults within an entry, and assign fault classes equally.

Several of the Defects4J patches combine the repair with refactoring (see \cref{listing:math46}).
This biases a repair classification and may inhibit our fault classification; we mitigate this by considering different possible repairs for a fault.

We consider only Defects4J projects, all of which are written in Java, as our benchmark set.
Our results may not necessarily extend to projects of different types or programming languages.
We leave this evaluation to future work.

\section{Dataset Statistics}\label{sec:ds}

\begin{figure}[!htb]
  \centering
  \includegraphics[width=5.5cm]{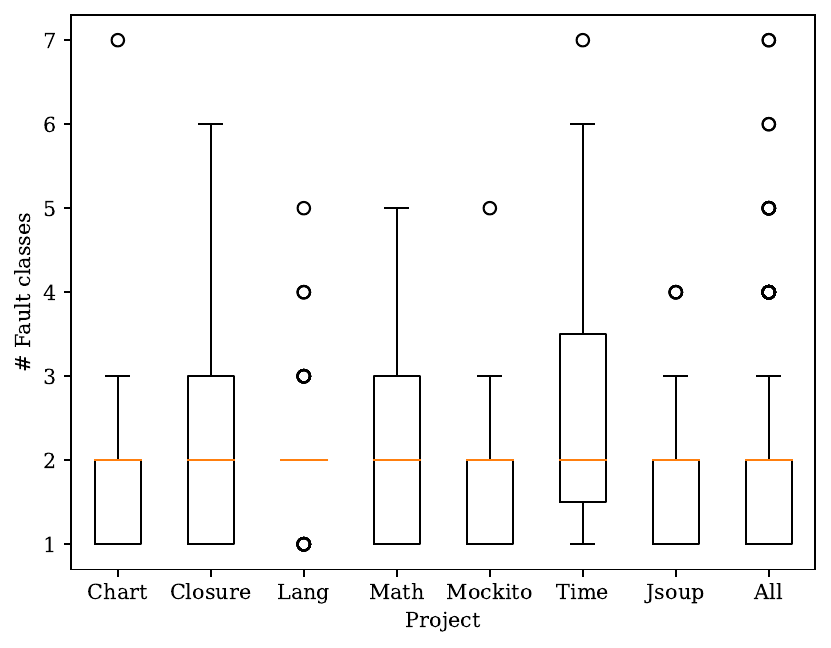}
  \caption{The distribution of number of fault classes assigned per fault, per project and overall}\label{fig:whiskers}
\end{figure}

\cref{fig:whiskers} shows, for each individual project and aggregated over all projects, the distribution of the number of fault classes assigned to the faults.
Most projects have a median (orange line) and upper quartile of two classes assigned to each fault.
Overall, the majority of faults are each assigned between one and three fault classes, but 
some faults in Chart and Time are assigned as many as seven of the eight possible fault classes.
All faults are assigned at least one fault class, as none of the faults were considered unclassifiable.

\cref{tab:project} shows the frequency of each fault class, in terms of the number of faults it is assigned to, over the 488~classified Defects4J faults.
We see that \emph{definition} fault (data-flow) is the most common fault class for each project.
On the other hand, \emph{unconditional statement order} fault (control-flow) is the least common class.
Overall, control-flow classifications outweigh data-flow classifications roughly by a factor of 1.5 to 1, even if the individual control-flow classes are less frequently assigned.
Because the fault classes are not disjoint, we consider disjoint fault types, where a fault is assigned only control-flow classes (pure CF), only data-flow classes (pure DF), or both (CF/DF). Pure CF faults outweigh pure DF faults roughly by a factor of 2 to 1, but the majority of faults display a mixture of control- and data-flow characteristics.

\begin{figure*}[!htb]
  \centering
  \begin{subfigure}[t]{0.49\textwidth}
    \centering
    \includegraphics[width=5.74cm]{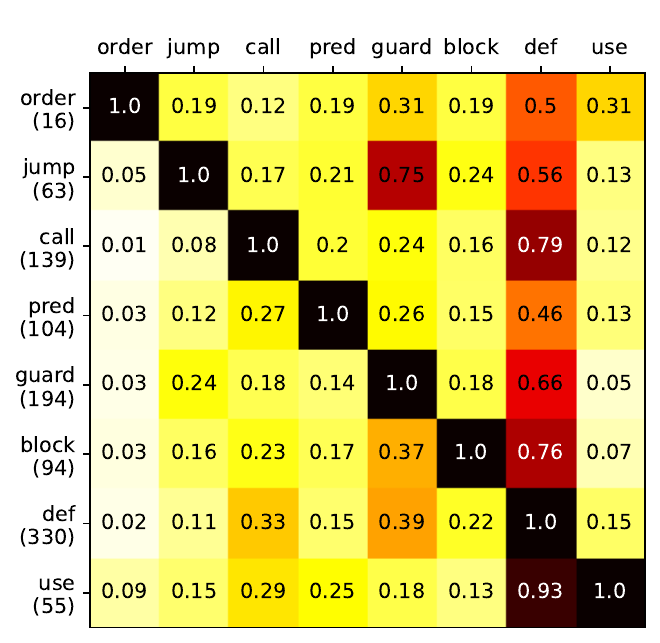}
    \vspace{-1mm}
    \caption{Fault class vs.\ fault class}\label{fig:confm_ff}
    \vspace{1mm}
  \end{subfigure}
  \begin{subfigure}[t]{0.49\textwidth}
    \centering
    \includegraphics[width=5.265cm]{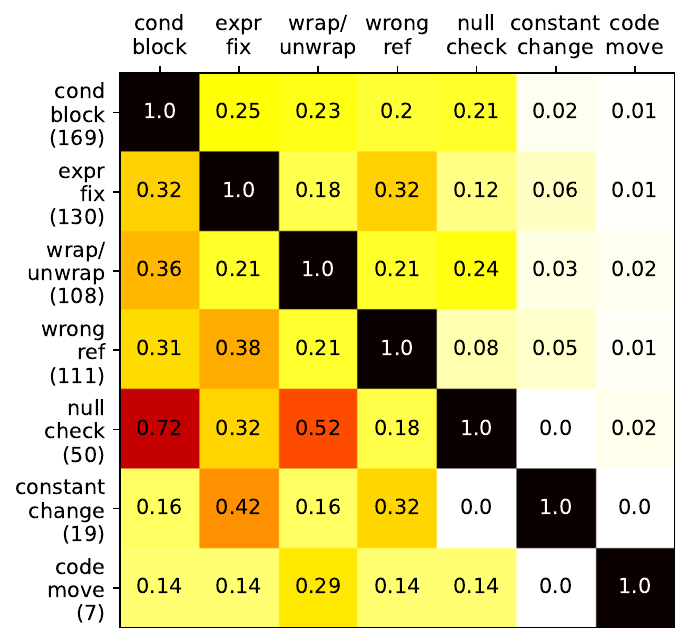}
    \vspace{-1mm}
    \caption{Repair class vs.\ repair class}\label{fig:confm_rr}
  \end{subfigure}
  \begin{subfigure}[t]{0.49\textwidth}
    \centering
    \includegraphics[width=5.94cm]{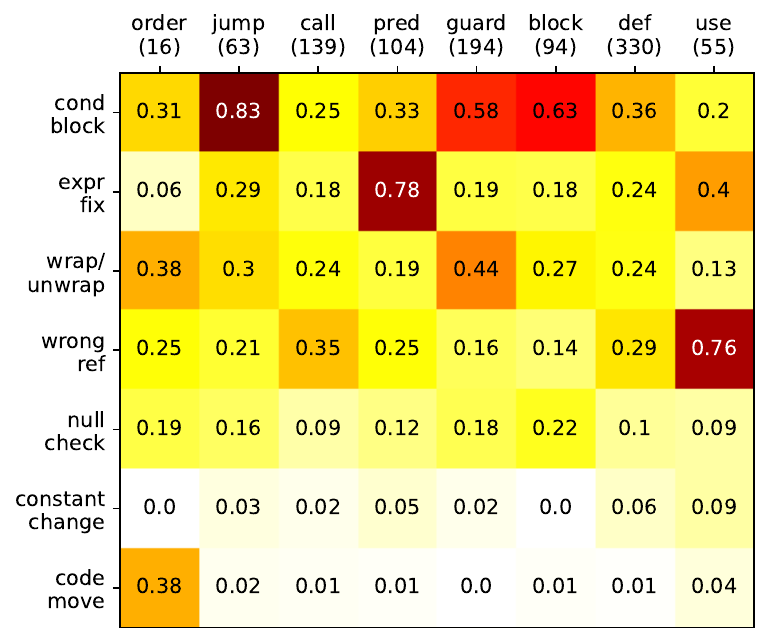}
    \caption{Repair class vs.\ fault class (\% of fault class)}\label{fig:confm_f}
  \end{subfigure}
  \begin{subfigure}[t]{0.49\textwidth}
    \centering
    \includegraphics[width=5.94cm]{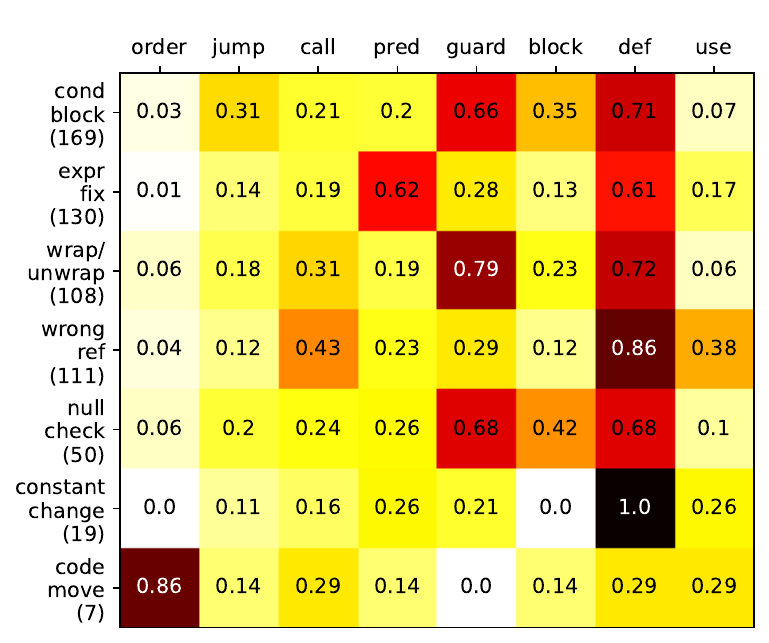}
    \caption{Repair class vs.\ fault class (\% of repair class)}\label{fig:confm_r}
  \end{subfigure}
  \caption{Co-occurrence matrices between classes (our fault classes, described in \S\ref{sec:fgbc}; repair classes~\cite{Defects4JDissection});\\ two classes co-occur for a fault if and only if both classes are assigned to the fault}\label{fig:confm}
\end{figure*}

\cref{fig:confm_ff} shows the details of the co-occurrences between our fault classes over the full dataset.
Each cell $i,j$ shows the percentage of faults assigned fault class $i$ that are also assigned fault class $j$.
There is high co-occurrence between several classes,
especially with \emph{definition} fault.
Due to the discussed distribution, faults are often assigned more than one class, and definition fault is the most frequent class.

Sobreira \emph{et al.}~\cite{Defects4JDissection} investigate 395 of these faults; however, they focus on the repair actions and patterns, rather than fault types.
The repair actions are very fine-grained, since they distinguish between addition, removal, and modification in nearly all syntactic categories.
This leads to 28~different classes.
The nine repair patterns are more abstract, and thus closer to our classification.
For example, the \emph{wraps-with/unwraps-from} pattern is comparable with our \emph{predicate existence} fault class (\ref{par:guard}).
However, some repair pattern categories are not useful for our purposes.
For example, \emph{single line} does not give much information about the fault itself---the fault is fixed in one line, but this is not the only possible fix.
\cref{fig:confm_rr} shows the co-occurrences of seven repair classes from~\cite{Defects4JDissection} over all projects.
These generally show lower co-occurrence than our own.

\cref{fig:confm_f} shows the percentage correlation between our assigned fault classes with the repair pattern classes~\cite{Defects4JDissection}, and \cref{fig:confm_r} shows the reverse.
We see several one-sided co-occurrences between the classifications, with
(\emph{cond block, guard}) and (\emph{expr fix, pred}) the only pairs to show two-sided majority co-occurrence.
The generally low co-occurrence illustrates that the two classifications are distinct.
\cref{fig:confm_ff,fig:confm_r} show that both fault and repair class assignments often correlate with the definition fault class,
implying that these faults and fixes often focus on definitions.

\section{Data Availability}\label{sec:da}
The classification results are available at \url{https://doi.org/10.5281/zenodo.15032242}.
The replication package is available at \url{https://doi.org/10.6084/m9.figshare.28855634}.

\section{Dataset Usage}\label{sec:duafrq}

Our fault classification scheme is novel for its exploration of intrinsic \emph{fault} characteristics, not just repairs.
Studies that use our classified dataset as a benchmark can thus investigate the performance of novel debugging techniques in terms of the true nature of the faults.
The fault classes are generalizable to any programming language in which programs can be represented as control- and data-flow graphs, e.g., Python, and we plan to use our fault taxonomy to investigate the BugsInPy dataset~\cite{bugsinpy}.

We have already considered the research question ``Are spectrum-based fault localization (SBFL) techniques more effective for certain fault types?''
Our preliminary results show that SBFL performs worse on pure control-flow faults, likely because such faults often involve missing code, which is difficult to localize.
This is supported by Szatm\'{a}ri \emph{et al.}~\cite{Szatmari_Vancsics_Beszedes_2020}'s investigation of the effect of fault fix types on SBFL efficiency:
they find that adding an \emph{else}-branch (predicate block fault---control-flow) is significantly harder to localize, while changing a method call parameter (definition fault---data-flow) is significantly easier to localize than other fix types.

Our classification can be used as a benchmark dataset to validate SBFL techniques that are optimized for specific control- or data-flow fault types, or as a training set for a large language model (LLM) classifier.
We can also investigate the research question, ``Can the classification of an unknown fault be predicted from SBFL performance?''
In other words, if SBFL can localize a fault in the top-$N$ of the ranking, is it more likely to be a data-flow fault?

Other possible applications include automated program repair. 
For example, we can use the fault class frequency information to guide automated program repair to attempt fixes in order of fault class frequency; 
we could even tailor individual repair operations to the specific fault classes.

\section{Related Work}\label{sec:rw}
Goodenough and Gerhart~\cite{Goodenough_Gerhart_1975} give a phase-based error
classification (\emph{construction}, \emph{specification}, \emph{design},
\emph{requirements}), but this is too abstract for any detailed code analysis. 
They also classify program errors into \emph{missing control-flow paths},
\emph{inappropriate path selection}, and \emph{inappropriate or missing
action}.  These classes are similar to some of our fault classes, 
but the classification does not account for all faults found in practice.
Duncan and Robson~\cite{Duncan_Robson_1996} focus on faults in C programs.
Their main classes are \emph{problematic construct}, \emph{multiple file usage}, and \emph{portability issues}.
However, some of the faults are overly specific (e.g., \emph{confusion of \& and \&\&}) or not applicable to our dataset (e.g., \emph{scanf with wrong types}).

Pan \emph{et al.}~\cite{Pan_Kim_Whitehead_2009} describe a classification
scheme of 27 bug fix patterns.  The scheme is quite fine-grained, e.g.,
\emph{change of if condition expression} and \emph{change of loop predicate}
form different classes, despite both being predicate faults.  However, since
certain fault characteristics can be deduced from the fix patterns, the schema
has found use and adoption.
Nath \emph{et al.}~\cite{Nath_Merkel_Lau_2012} use these classes as a fault
classification scheme for Java, while Lucia \emph{et
al.}~\cite{Lucia_Lo_Jiang_Thung_Budi_2014} limit their scheme to eight
categories, six of which are taken from Pan \emph{et
al.}~\cite{Pan_Kim_Whitehead_2009}, with two categories added (incl.\ \emph{others}).  
However, predicate faults remain divided between conditionals and loops.
Szatm\'{a}ri \emph{et al.}~\cite{Szatmari_Vancsics_Beszedes_2020} investigate the effect of these bug fix patterns (specifically if- and sequence-related) on SBFL efficiency.
They show that certain patterns are significantly easier (e.g., \emph{change to an if condition}) or significantly harder (e.g., \emph{adding/removing an else branch}) to localize than others.
Thus localization efficiency can vary within a broader fix pattern, e.g., if-related fixes.

DiGiuseppe and Jones~\cite{DiGiuseppe2015} investigate fault localization
performance wrt.\ four fault taxonomies, of which three are quite coarse, with
only between four and five fault types.  The single fine-grained taxonomy,
taken from Hayes \emph{et al.}~\cite{Hayes_Chemannoor_Holbrook_2011}, is a
generic fault taxonomy and includes classes such as \emph{visual stimulation}
and \emph{documentation}, which are typically not useful in bug datasets.
DiGiuseppe and Jones find no correlation between their classes and
localization effectiveness; however, our preliminary results show that such a
correlation likely exists for our classification, indicating it is more
suitable for such applications.
Catolino \emph{et al.}~\cite{Catolino_Palomba_Zaidman_Ferrucci_2019} analyze the bug reports of $119$~GitHub projects and form a taxonomy of nine high-level fault classes: \emph{configuration, network, database-related, GUI-related, performance, permission/deprecation, security, program anomaly, test code-related}.
Humbatova \emph{et al.}~\cite{Humbatova_Jahangirova_Bavota_Riccio_Stocco_Tonella_2020} investigate the classification of deep learning faults, forming a taxonomy of five top-level categories (\emph{model, tensors \& input, training, GPU usage, API}), with \emph{model, tensors \& input, training} each having multiple subcategories.

In recent years, researchers have investigated the use of deep learning models in automated software fault classification.
Li \emph{et al.}~\cite{Li_Shi_Duan_Liu_Yang_2024} use a pre-trained BERT model to categorize faults, and show that their model improves on existing models.
However, their fault classes (\emph{hardware, algorithm, software, human factors, interaction, network}) are rather abstract and unsuited for fault localization and repair benchmark datasets, which generally consist of software faults.
Furthermore, the automated fault classification was tested on ``detailed case descriptions''~\cite{Li_Shi_Duan_Liu_Yang_2024}, which are not always available in fault datasets, and require a programmer's understanding of the fault in question.
This type of classification may not capture the full nuance of the fault.


\section{Conclusion and Future Work}\label{sec:conclusion}
In this paper, we introduce a novel fault classification scheme based on program flow graphs.
We classify 488 faults from seven real-world projects in the Defects4J repository.
The distribution of the number of fault classes assigned per fault is generally similar between different projects; the average is around two classes assigned per fault, but faults are generally assigned between one and three classes.
The fault class frequency over all faults shows that the definition fault (data-flow) is the most frequent fault class with 330~faults classified as such.
The unconditional statement order fault (control-flow) is the least frequent, with only 16~faults classified as such.
However, pure data-flow faults are the least common in the dataset ($16.6\%$), compared with pure control-flow ($31.6\%$) and mixed control- and data-flow ($51.8\%$) faults.

In this paper we report intermediate results from ongoing work. 
Current work includes a more expansive classification, performed by a larger group of raters on larger-scale datasets and different programming languages (e.g., BugsInPy~\cite{bugsinpy}), using a more detailed categorization.
We also plan to consider dynamic flow graphs and to test scalability and statistical significance among the fault classes.
We can further investigate the effect of different fault classes on fault localization and repair ability. 
Finally, automated analysis of the graph structure may be possible, but the use of LLMs or deep neural networks (DNNs) for automatic fault classification and prediction is also worth considering.

\bibliographystyle{ACM-Reference-Format}
\bibliography{refs}

\end{document}